# Dual functions of anti-reflectance and surface passivation of the atomic layer deposited $Al_2O_3$ films on crystalline silicon substrates

Li Qiang Zhu[a)], Xiang Li, Zhong Hui Yan, Hong Liang Zhang, and Qing Wan[b)]

Ningbo Institute of Materials Technology and Engineering, Chinese Academy of Sciences, Ningbo 315201, People's Republic of China

Surface anti-reflectance and passivation properties of the $Al_2O_3$ films deposited on crystalline Si substrates by atomic layer deposition are investigated. Textured Si with 100 nm $Al_2O_3$ shows a very low average reflectance of ~2.8 %. Both p-type and n-type Si wafers are well passivated by $Al_2O_3$ films. The maximal minority carrier lifetimes are improved from ~10 μs before $Al_2O_3$ passivation to above 3 ms for both p-type and n-type Si after $Al_2O_3$ passivation layer deposition and annealing at an appropriate temperature. Our results indicate the dual functions of anti-reflectance and surface passivation in c-Si solar cell applications.

---

[a)] E-mail address: lqzhu@nimte.ac.cn
[b)] Corresponding author. E-mail address: wanqing@nimte.ac.cn

$Al_2O_3$ thin films are of interest for various applications such as high-k gate candidates [1], protective coatings for a wide range of chemicals [2], optical waveguides [3], etc. Recently, $Al_2O_3$ have been received tremendous interest in the photovoltaic community due to their effective surface passivation to improve the efficiency. Ultra thin $Al_2O_3$ layers deposited on photoanodes used in dye-sensitized solar cells prevents the back recombination between electrons and the photoanodes, resulting in the improved cell efficiency. [4] Surface passivation of CIGS with atomic layer deposited (ALD) $Al_2O_3$ is also demonstrated due to the field shielding effects. [5] $Al_2O_3$ thin films provide excellent surface passivation on both lightly and highly doped p- and n-type crystalline silicon surfaces, resulting in the improved efficiency [6-10], which is due to the low density of interface defects $D_{it}$ in the range of $10^{11} cm^{-2} eV^{-1}$ as well as a field effect passivation with a high density of fixed negative charges $Q_{fix}$ above $10^{12} cm^{-2}$. [6, 11] To active the passivation, an annealing step at moderate temperatures after deposition was reported to be essential.[12]

At the same time, effective anti-reflective plays an important role for solar cells efficiency improvement. Conventionally, textured Si surface coated with $SiN_x$ anti-reflective layer from production line shows an average reflectance of below 5%. It is reported [13] that the reflective index of $Al_2O_3$ thin films is ~1.6 at wavelength of 630nm, indicating that it is suitable for anti-reflectance applications in c-Si solar cells. Though the $Al_2O_3$ films have been widely employed to passivate the c-Si surface with a thickness below 50nm, the anti-reflective properties of $Al_2O_3$ films have not been reported yet. In this letter, both the anti-reflectance properties and the passivation properties of the thermal ALD $Al_2O_3$ have been studied. A minimal reflectance of ~2.8 % was addressed. A maximal minority carrier lifetime of 4.7 ms and 3.4 ms were obtained for $Al_2O_3$ passivated p-Si and n-Si wafers, respectively. Such results indicate the dual functions of anti-reflectance and surface passivation in c-Si solar cell applications.

Commercially available single crystalline, 380μm-thick (100)-oriented p-type Czochralski (CZ) silicon wafers and 400μm-thick (100)-oriented n-type Czochralski (CZ) silicon were used as the substrates. For the anti-reflective testing, the wafers were textured by NaOH solution. Then the substrates underwent an HCl solution dip for 5min followed by a dilute HF dip. A de-ioned water rinse was adopted after each chemical dip steps. The $Al_2O_3$ films were deposited on the textured wafer for anti-reflectance studies by a thermal NCD 200B ALD reactor. To test lifetime, an $Al_2O_3$ film was deposited on both sides of the shiny-etched wafer to obtain symmetric lifetime sample. The stoichiometric $Al_2O_3$ films were deposited at 200℃ with a 100sccm background flow of $N_2$. A cycle in the reactor consisted of a 0.3s injection of $Al(CH_3)_3$ vapors followed by 7s $N_2$ purge. The oxidation step consisted of a 0.1s injection of $H_2O$ vapor followed by a 7s purge with $N_2$ resulting in a deposition rate of 1.25 Å/cycle. 800, 560, and 240 ALD cycles were adopted on the textured p-type CZ silicon wafers for anti-reflectivity testing. While 800 and 240 ALD cycles were adopted on the CZ silicon wafers for lifetime testing. Control samples are also prepared on the shiny-etched p-type CZ silicon wafers received HF-last rinse to measure the film thickness by spectroscopy elliposometry. The post-deposition annealing was performed in a quartz furnace at different temperature in atmosphere ambient.

The total spectral reflectance was measured by AudioDev's Helios LAB-rc system. Reflected light from a broadband halogen light source is collected and detected by a special designed integrating sphere. Therefore, all reflected or scattered light will be measured. Microwave photoconductance decay lifetime measurements were carried out to determine the passivation quality on a Semilab WT-2000PVN lifetime tester with excess carriers generated by a 200 ns laser pulse at a wavelength of 904 nm and a spot size of 1mm$^2$. The effective minority carrier lifetime $\tau_{eff}$ depends on both the bulk minority carrier lifetime $\tau_{bulk}$ and the surface recombination velocity $S_{eff}$.[14]

$$\frac{1}{\tau_{eff}} = \frac{1}{\tau_{bulk}} + \frac{2S_{eff}}{W}$$

where $S_{eff}$ is the surface recombination velocity and W is the wafer thickness. $S_{eff}$ is calculated from the effective minority carrier lifetime of the samples, considering their intrinsic lifetime values. The bulk minority carrier lifetime was assumed to be infinite. Accordingly, the calculated $S_{eff}$ value marks an upper limit to the effective surface recombination velocity.

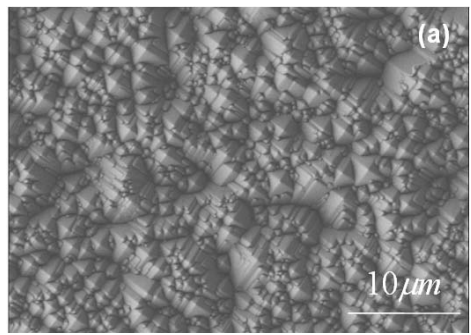

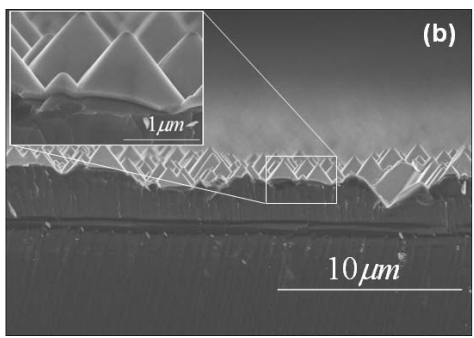

Fig.1 (a) SEM top view image of the $Al_2O_3$ coated textured Si surface of a (100)-oriented monocrystalline Si wafer after anisotropical wet chemical etching leaving (111) planes. (b) The micrograph of the cross-section of the $Al_2O_3$ coated textured Si.

A wet chemical alkaline anisotropically etching solution leads to the pyramidal surface topography. Fig.1 (a) shows the scanning electron microscopy (SEM) images of the surface morphology of the $Al_2O_3$ coated textured Si surface. (111) oriented Si crystal planes remain at the surface forming the pyramids. Fig.1 (b) shows a SEM cross section of a portion of the $Al_2O_3$ coated texture Si wafer. The thickness of the textured surface is estimated to be ~3 μm. An $Al_2O_3$ layer is obviously observed to be covered on the pyramids with the thickness of ~100 nm. Such textured surface helps to reduce the optical reflectance from 29.4% before texturing to 14.2% after texturing in average. While the $Al_2O_3$ over layer reduce the optical reflectance further, as will be discussed below.

Spectroscopic ellipsometry (SE) has been employed to investigate the optical characteristics of $Al_2O_3$ films. The thicknesses are determined to be ~100 nm, ~70 nm and ~30nm for 800, 560, and 240 ALD cycles, respectively. Fig.2 (a) presents the refractive index (*n*) and extinction coefficients (*k*) of the as-deposited 100 nm $Al_2O_3$ in the wavelength range of 190 nm-1400 nm. The refractive index is measured to be ~1.65 at wavelength of 630 nm, similar to the reported value.[13] It is interesting to note that the extinction coefficient is determined to be close to zero for wavelength above 200nm. The absorption coefficients (*α*) could be obtained by a relationship, $α=4πk/λ$, where λ was the wavelength of a photon as shown in Fig.2 (b). A significant increase of the absorption coefficient *α* at higher photon energy could be attributed to the band–band transitions. Since $Al_2O_3$ systems have an indirect fundamental gap, the interband absorption can be expressed by the following equation:[15]

$$\alpha h\upsilon \propto (h\upsilon - E_g)^2$$

where $E_g$ is optical band gap energy. The inset in Fig.2 (b) illustrates the $(\alpha h\nu)^{1/2}$ vs photon energy (hν) curves. The linear behavior supports the interband transition. The extracted optical band gap ($E_g$) is ~6eV, which means that the deposited $Al_2O_3$ layer is transparent for the wavelength above 200 nm, ie, no absorption loss in the $Al_2O_3$ layer occurs for wavelength above 200 nm. The obtained results are meaningful for the anti-reflectance applications in solar cells.

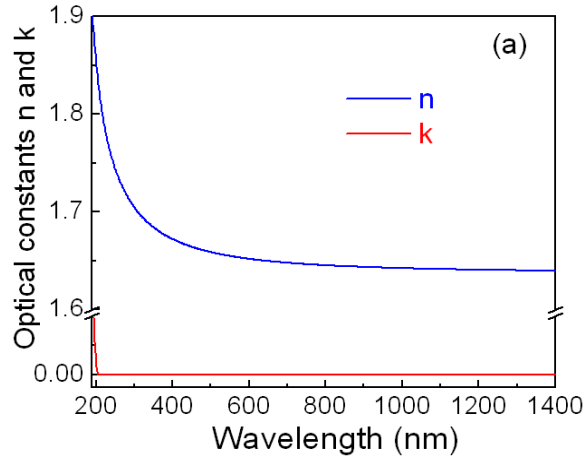

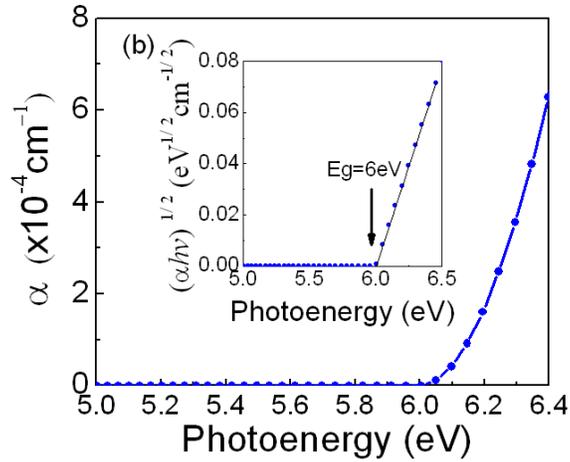

Fig.2 (a) Optical constants n and k for the as-deposited ~100 nm $Al_2O_3$ as a function of the wavelength obtained with spectroscopic ellipsometry measurements. (b) α vs *hv* plot of as-deposited 100 nm $Al_2O_3$ films. The inset figure shows the $(\alpha h v)^{1/2}$ vs *hv* plot. The intersection indicates the $E_g$ value.

Fig.3 (a) illustrates the photographs of the textured Si coated with 100 nm, 70 nm and 30 nm $Al_2O_3$, showing the best anti-reflectance properties for 100 nm $Al_2O_3$ coated textured Si. Photographs for textured Si and $SiN_x$ coated textured Si are included as references. Fig.3 (b) shows the reflectance spectrum of the $Al_2O_3$ coated textured Si wafers with different $Al_2O_3$ thickness. Reflectances for the textured Si and the $SiN_x$ coated textured Si are also included for comparison. The standard $SiN_x$ coated textured Si are obtained from the production line, noted as standard (STD) sample. The average reflectance is ~14.2 % for the textured Si. It decreases to ~10.6 % and ~4.2 % when depositing ~30 nm $Al_2O_3$ and ~70 nm $Al_2O_3$, respectively. Meaningfully, the reflectance spectra for ~100 nm $Al_2O_3$ coated textured Si are quite similar to that of the STD sample. At the low wavelength, the reflectance for ~100 nm

Al$_2$O$_3$ is a little higher than the STD sample, while at the higher wavelength, the reflectance is a little lower than the STD sample, which results in the best reflectance of ~2.8 % for ~100 nm Al$_2$O$_3$ on the textured Si, close to 2.9 % for the STD sample. The results indicate the potential anti-reflectance applications of Al$_2$O$_3$ in c-Si solar cells.

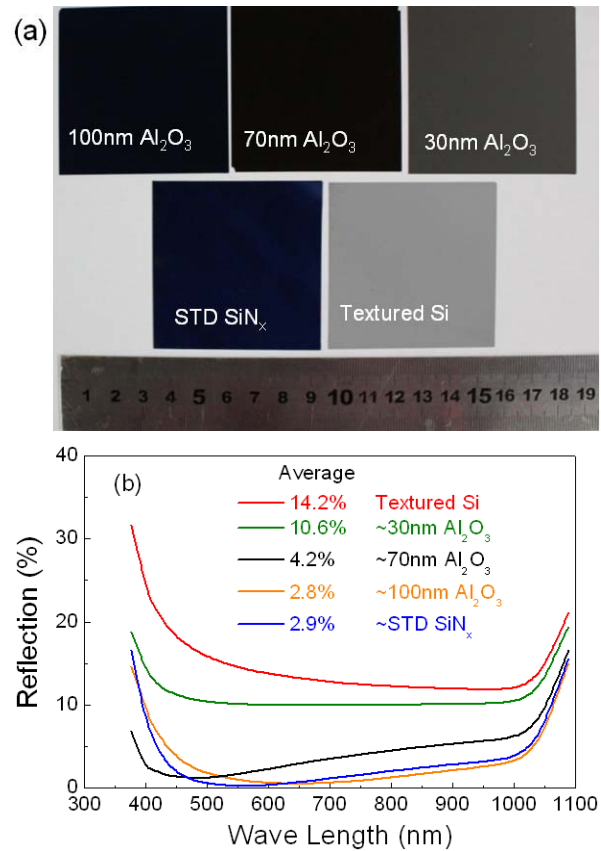

Fig.3 (a) A photograph of the texture Si coated with 100nm, 70nm, 30nm Al$_2$O$_3$ and STD SiN$_x$. Textured Si as the reference. (b) Reflectance curves for the Al$_2$O$_3$ coated textured Si. Results for the textured Si and standard SiN$_x$ coated textured Si (STD SiN$_x$) are included for comparison.

Independent of the deposition method, a post-deposition anneal is necessary to activate the passivating properties of $Al_2O_3$. Fig.4 illustrates the effective carrier lifetime (Fig.4 (a)) and the surface recombination velocity (Fig.4 (b)) of p-type CZ Si wafers as a function of the applied thermal treatment for 30 nm and 100 nm $Al_2O_3$ coated Si. For the original Si wafer, a low lifetime of ~6 μs is obtained. After depositing a 100 nm-thick $Al_2O_3$ layers, a moderate surface passivation level has been addressed, yielding an effective lifetimes $\tau_{eff}$ of ~140 μs, similar to what has been observed for conventional thermal ALD.[16] While depositing a 30 nm-thick $Al_2O_3$ layers results in a high effective lifetimes $\tau_{eff}$ of ~910 μs. To study the full potential for the surface passivation and the thermal stability of the $Al_2O_3$ layers deposited in this work, the lifetime samples were exposed to a post-deposition annealing in atmosphere ambient for 5 min with temperature ranging from 300 ℃ to 650 ℃. A flash annealing was also performed at 900 ℃ for 3 s. For the 100 nm $Al_2O_3$ coated lifetime samples, annealing performed at 600 ℃ for 5 min yields a good passivation with the lifetime of ~750 μs. While for the 30nm $Al_2O_3$ coated lifetime samples, a best passivation is obtained at 350 ℃ with a lifetime of ~4.7 ms. Annealing the sample at higher temperature results in the deteriorated lifetime. The flash annealing for 30nm $Al_2O_3$ coated lifetime samples at 900 ℃ for 3 s yields a moderate level of surface passivation with lifetime of 120 μs.

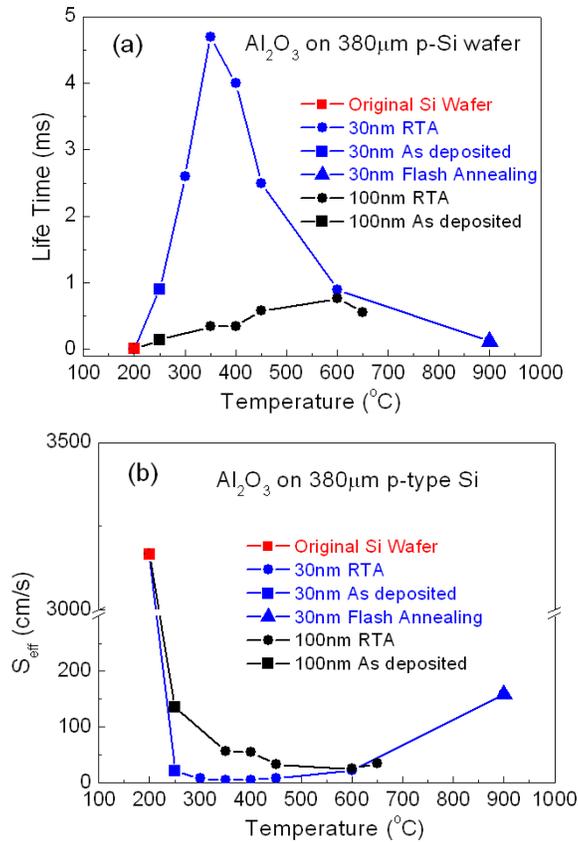

Fig.4 (a) Life time values for p-type Si passivated by 100 nm and 30 nm $Al_2O_3$. (b) Effective surface recombination velocity ($S_{eff}$).

The upper limit of the effective surface recombination velocity was also calculated as shown in Fig.4 (b). For the original Si wafer, a high $S_{eff}$ value of ~3170 cm/s is determined. The deposition of 100 nm $Al_2O_3$ layer results in a moderate $S_{eff}$ of ~130 cm/s, while the deposition of 30nm $Al_2O_3$ results in a low $S_{eff}$ of ~20 cm/s. The post-deposition annealing treatment results in the improved $S_{eff}$. For 100 nm $Al_2O_3$ coated Si wafers, the best results are obtained at 600 ℃, yielding a lowest $S_{eff}$ of ~25cm/s. While for 30 nm $Al_2O_3$ coated Si wafers, a very low $S_{eff}$ <20 cm/s is obtained. The lowest $S_{eff}$ of ~4 cm/s is addressed at 350 ℃. The flash annealing at 900 ℃ for 3s yields a moderate level of surface passivation with $S_{eff}$ of 160 cm/s, respectively.

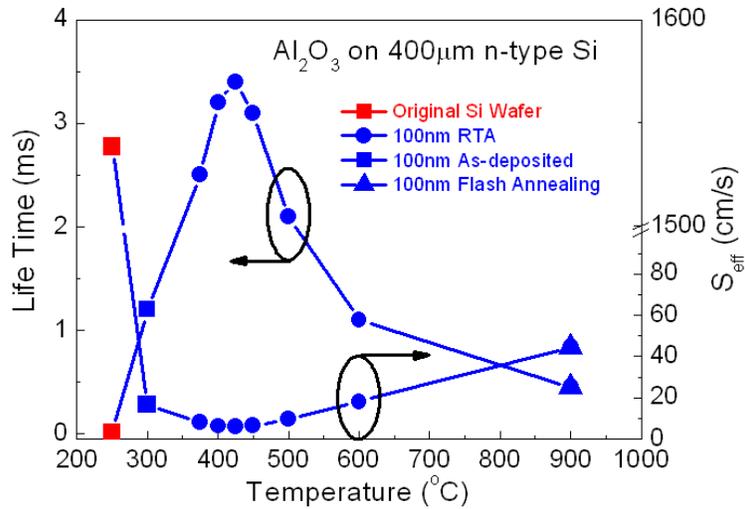

Fig.5 Life time values for n-type Si passivated by 100 nm $Al_2O_3$ and the effective surface recombination velocity ($S_{eff}$).

Similarly, n-type Si wafers are also passivated by 100 nm thick $Al_2O_3$, as shown in Fig.5. For the original Si wafer, a low lifetime of ~13 μs is obtained. After passivating with a 100 nm thick $Al_2O_3$, the effective minority carrier lifetime increased to 1.2 ms and 3.4 ms for the as-deposited sample and the 425 ℃ annealed sample, repectively. The flash annealing at 900 ℃ for 3 s yields a moderate level of surface passivation with lifetime of 450 μs. The upper limit of the effective surface recombination velocity is obtained. For the original Si wafer, a high $S_{eff}$ of ~1540 cm/s is determined. The deposition of 100 nm $Al_2O_3$ layer results in a low $S_{eff}$ of ~16 cm/s. The lowest $S_{eff}$ of ~6 cm/s is addressed at 425 ℃. The flash annealing at 900 ℃ for 3s yields a moderate level of surface passivation with $S_{eff}$ of 44 cm/s, respectively.

The activation of the passivation are attributed to a strong field effect passivation caused by a high negative fixed charge density in the $Al_2O_3$ films located close to the interface.[11] While the deteriorating of the passivation at the higher temperature would be attributed to the deteriorated interface properties, ie, the increased $D_{it}$ and the defects density.

In summary, $Al_2O_3$ layers were deposited by thermal ALD on p-type and n-type CZ Si wafers. The textured Si coated with 100nm $Al_2O_3$ shows a low average reflectance of ~2.8 %. Both p-type and n-type Si wafers are well passivated by $Al_2O_3$ films. The maximal minority carrier lifetimes are improved from ~10 μs before $Al_2O_3$ passivation to above 3 ms for both p-type and n-type Si after $Al_2O_3$ passivation and annealing. Our results indicate the dual functions of anti-reflectance and surface passivation in c-Si solar cell applications.

**Acknowledgement**

Programs supported by Ningbo Natural Science Foundation (2011A610202), and the National Natural Science Foundation of China (11104288).